\documentclass[12pt,a4paper]{article}
\usepackage{amsmath}
\usepackage{amssymb}

\allowdisplaybreaks

\begin{document}

\title{Warped embeddings between \\
Einstein manifolds}

\author{Huan-Xiong Yang\\
Interdisciplinary Center for Theoretical Study,\\ 
University of Science and Technology of China, Hefei, 230026, China\\
{}\\
Liu Zhao\\
School of Physics, \\
Nankai University, Tianjin 300071, China}

\date{}

\maketitle

\begin{abstract}
Warped embeddings from a lower dimensional Einstein manifold into a higher dimensional one are analyzed. Explicit solutions for the embedding metrics are obtained for all cases of codimension 1 embeddings and some of the codimension $n>1$ cases. Some of the interesting features of the embedding metrics are pointed out and potential applications of the embeddings are discussed.
\end{abstract}

\section{Introduction}

Warped geometries have attracted considerable attentions from diverse areas of modern theoretical physics ranging e.g. from braneworld models to warped compactification of string theory. Most studies on warped geometries are based on a specific ansatz on the warp factor, which may be subject to some additional discrete symmetry (e.g. in braneworld theories the discrete symmetry is often taken to be $Z_2$), and the number of warped extra dimensions is often limited to 1.

Meanwhile, Einstein manifolds are a class of spacetime manifolds which play a very important role in General Relativity. Such spacetime manifolds correspond to vacuum solutions of Einstein's theory of gravity either with or without a cosmological constant. In many cases, black hole and cosmological solutions are either a specific example of or intimately related to this class of spacetimes. Under the assumption that extra spacetime dimensions exist it is interesting to ask whether an Einstein manifold of lower dimension can be embedded into another Einstein manifold of higher dimension.

In this article, we shall consider the problem of embeddings between Einstein manifolds of different spacetime dimensions. From the string theory or experimental detection perspectives, there is no reason to restrict the number of extra dimensions to 1, and no additional discrete symmetry is required {\it a priori}. To be more concrete, what we shall study is the the following $D$-dimensional spacetime (bulk) with metric $\hat{g}_{MN}$ having a warped form
\begin{equation}
ds^{2}=\hat{g}_{MN}dX^{M}dX^{N}=e^{a(y)}g_{\mu \nu}(x)dx^{\mu}dx^{\nu}%
+h_{ij}(y)dy^{i}dy^{j}, \label{eq:1}
\end{equation}
where $X^{M}=(x^{\mu},y^{i})$ are bulk coordinates, $a(y)$ is a scalar function of $y^{i}$ but not of $x^{\mu}$, 
with $M=0,1,...,D-1$, $\mu=0,1,...,d-1$ and $i=d,...,D-1$. Quantities associated with the bulk are labeled by a hat.  The difference $n=D-d$ between the dimensions of the bulk and the submanifold with metric $g_{\mu\nu}$ will be referred to as the codimension of the embedding.

The metric $\hat{g}_{MN}$ of the bulk and 
$g_{\mu \nu}$ of the $d$-dimensional submanifold are both required to be Einstein manifolds, i.e.
\begin{align}
\hat{R}_{MN} = \frac{2\hat{\Lambda}}{D-2} \hat{g}_{MN} ,\quad 
R_{\mu\nu} &= \frac{2\Lambda}{d-2} g_{\mu\nu} .\label{r-eq}
\end{align}
Notice that simply identifying a cosmological constant is not sufficient to justify the spacetime to be a de Sitter/anti-de Sitter or Minkowski spacetime, because in the usual notion of de Sitter/anti-de Sitter and Minkowski spacetimes a maximal symmetry is implicitly assumed. However, in order to identify the nature of the bulk and the submanifold we are going to deal with, we shall abuse terminologies and denote Einstein manifolds with negative, zero and positive cosmological constants respectively with $AdS_{D}$, $M_{D}$ and $dS_{D}$, leaving the maximal symmetry ansatz aside\footnote{The spacetime manifolds that are referred to as of Minkowski in this paper may be properly called Ricci-flat Einstein manifolds.}. 

Thus we are facing up with 9 different types of embeddings, i.e.
\begin{equation}
AdS_{d} \subset \left\{
\begin{array}{l}
AdS_{D} \\
M_{D} \\
dS_{D}
\end{array}
\right. ,\qquad
M_{d} \subset \left\{
\begin{array}{l}
AdS_{D} \\
M_{D} \\
dS_{D}
\end{array}
\right. ,\qquad
dS_{d} \subset \left\{
\begin{array}{l}
AdS_{D} \\
M_{D} \\
dS_{D}
\end{array}
\right. . \label{9cases}
\end{equation}
The question to be asked can be clearly described as: for each of the 9 cases listed in (\ref{9cases}), can we find exact, analytic solutions of the embedding metric, especially, can we find $e^{a(y)}$ and $h_{ij}(y)$ for each cases without specifying the details of  $g_{\mu\nu}$? Partial answer to this question will be given in the main context of this article, and for each cases for which we can give an explicit answer we shall point out some of its intriguing features and make brief discussions on the potential applications of the specific answer.

By straightforward calculations we can see that 
the Ricci tensors for the bulk and
the submanifold are linked as follows:
\begin{align}
\hat{R}_{\mu \nu}  &  =R_{\mu \nu}(g)
+\frac{1}{2}\left[e^{-a}  \square ^{(h)} e^a +\frac{1}{2}(d-2) h^{ij}\nabla^{(h)}_i a \nabla^{(h)}_j a\right] \hat{g}_{\mu \nu} ,
\label{munu-D}\\
\hat{R}_{ij} &=R_{ij}(h)+\frac{d}{2}\left[ 
\nabla^{(h)}_i \nabla^{(h)}_j a
 +\frac{1}{2}\nabla^{(h)}_i a \nabla^{(h)}_j a \right], \label{ij-D}
\end{align}%
where $\square^{(h)}$ and $\nabla^{(h)}_i$ respectively denote the D'Alembertian and covariant derivatives associated to 
the metric $h_{ij}$.  
Similar calculations can also be found in \cite{Leblond}, but with different notations. 

It is better to parametrize the cosmological constants $\hat{\Lambda}$ and $\Lambda$ in a suitable way so that the solutions look more elegant. We choose the parametrization
\begin{align}
\hat{\Lambda}= \mathrm{sign}(\hat{\Lambda})\frac{(D-1)(D-2)}{2} \hat{k}^2, \quad
\Lambda =  \mathrm{sign}({\Lambda})\frac{(d-1)(d-2)}{2} k^2,  \label{lambda}
\end{align}
where $\mathrm{sign}(\hat{\Lambda})$ denotes the signature of 
the cosmological constant $\hat{\Lambda}$ (which takes one of 
the ``values'' $(-, 0, +)$), and $\hat{k}$ and $k$ are some
positive numbers. In section 4, we also write 
\begin{align}
\hat{\kappa}^2= (D-1) \hat{k}^2, \quad
\kappa^2 = (d-1)k^2.  \label{lambda-2}
\end{align}

\section{Codimension 1 cases} \label{S2}

For $n=1$ we can fix $h_{D-1,D-1}=-1$ without loss of generality. These are the most familiar cases because all braneworld models with codimension 1 belong to this class. Some of the 9 types of embeddings were already known from the studies of braneworld theories \cite{Pope}, and we include them here both for completeness and for triggering the ansatz for the higher codimension cases.

Inserting equations (\ref{munu-D}) and (\ref{ij-D}) into (\ref{r-eq}) and taking 
the parametrization (\ref{lambda}), we get the following equations for $a(y)$,
\begin{align}
&\frac{1}{2}\left[  a^{\prime \prime}(y)+\frac{D-1}{2}a^{\prime}(y)^{2}\right]
-\mathrm{sign}(\Lambda) (D-2) k^2e^{-a(y)} 
+\mathrm{sign}(\hat{\Lambda}) (D-1) \hat{k}^2 =0 , \label{embed-1}\\
&\frac{1}{2}\left[  a^{\prime \prime}(y)+\frac{1}{2}a^{\prime}(y)^{2}\right] 
+\mathrm{sign}(\hat{\Lambda}) \hat{k}^2 = 0. \label{embed-2}
\end{align}
Notice that (\ref{embed-1}) and (\ref{embed-2}) is a system of two equations for a single unknown function $a(y)$, so
the existence of solutions is not guaranteed. 
Different choices of $\mathrm{sign}(\hat{\Lambda})$ and $\mathrm{sign}({\Lambda})$ correspond to different types of embeddings mentioned in (\ref{9cases}), and the corresponding solutions varies drastically. 
We list the results here in a case by case manner.

\begin{enumerate}
\item {$AdS_{D-1}\subset AdS_{D}$}. 
In this case both $\hat{\Lambda}$ and $\Lambda$ are negative. The common solution to (\ref{embed-1}) and (\ref{embed-2}) is given by 
\begin{equation}
\exp(a(y))=\left(\frac{k}{\hat{k}}\right)^2 \cosh^2(\hat{k}y).
\label{C1}
\end{equation} 
Here and below we shall always absorb the unimportant integration constant by a shift in the coordinate $y$. 

\item {$AdS_{D-1}\subset M_{D}$}. 
Now we have $\hat{\Lambda}=0$ and $\Lambda < 0$. The solution reads
\begin{equation}
\exp(a(y))=-\left(ky\right)^2. \label{ads-m}
\end{equation}
The minus sign on the right hand side of (\ref{ads-m}) is unusual. It implies flipping the roles of spacelike and timelike coordinates in $x^\mu$. So, if originally there were only one timelike coordinate in $x^\mu$, the minus sign would results in a metric with multiple timelike coordinates unless $D=3$. 

\item {$AdS_{D-1}\subset dS_{D}$}, i.e.
$\mathrm{sign}(\hat{\Lambda})=+1$ and $\mathrm{sign}({\Lambda})=-1$. The solution is
\begin{equation}
\exp(a(y))=-\left(\frac{k}{\hat{k}}\right)^2\cos^2(\hat{k}y).
\end{equation}
Like in the previous case, this solution involves a bulk geometry with more than one timelike coordinates unless $D=3$.

\item {$M_{D-1}\subset AdS_{D}$}, i.e. $\hat{\Lambda}<0$ and $\Lambda=0$. The solution reads
\begin{equation}
\exp(a(y))=\exp(\pm 2\hat{k}y).
\end{equation}
This is the case on which the original Randall-Sundrum (RS) braneworld scenario \cite{Randall1,Randall2} was built. The difference between the RS braneworld model and the present embedding lies in that RS assumed an extra $Z_2$ symmetry which brought about a $\delta$-function-like discontinuity in the second derivatives of the metric, which is explained as the contribution from brane tension.

\item {$M_{D-1}\subset M_{D}$}.
Now both $\hat{\Lambda}$ and $\Lambda$ are zero. This is equivalent to saying that both 
$\hat{k}$ and $k$ are zero. The solution reads
\begin{equation}
a(y)=const.
\end{equation}
This is perhaps the most uninteresting of all possible cases, because it implies no warping in the bulk geometry at all.

\item {$M_{D-1}\subset dS_{D}$}. 
Now we have $\hat{\Lambda}>0$ and $\Lambda=0$. The only common solution to  (\ref{embed-1}) and (\ref{embed-2}) is essentially complex, 
\begin{equation}
\exp(a(y))=\exp(\pm 2i\hat{k} y),
\end{equation}
implying that the embedding $M_{D-1}\subset dS_{D}$ is impossible in principle if the bulk is a real manifold. This explains why there is no RS like braneworld model with a de Sitter bulk.

\item {$dS_{D-1}\subset AdS_{D}$}:
Here we have $\hat{\Lambda}<0$ and $\Lambda>0$. The solution is
\begin{equation}
\exp(a(y))=-\left(\frac{k}{\hat{k}}\right)^{2}\cosh^{2}(\hat{k}y).
\end{equation}
One sees that this type of embedding must also involve a bulk geometry with more than one timelike dimensions unless $D=3$.

\item {$dS_{D-1}\subset M_{D}$}, i.e. $\mathrm{sign}(\hat{\Lambda})=0$ and $\mathrm{sign}(\Lambda)=+1$. The solution
reads
\begin{equation}
\exp(a(y))=(ky)^{2}.
\end{equation}
This kind of embedding has already been used in the study of black rings and black strings \cite{Chu, Zhao}.

\item {$dS_{D-1}\subset dS_{D}$}, i.e. $\hat{\Lambda}>0$ and $\Lambda>0$. The corresponding solution is
\begin{equation}
\exp(a(y))=\left(\frac{k}{\hat{k}}\right)^{2} \cos^{2}(\hat{k}y).
\end{equation}
This is another interesting case which may found applications in the future.
\end{enumerate}

To summarize, not all of the 9 different types of embeddings yield physically interesting solutions. In some case the embedding is even impossible for real bulk manifold. We arrange the 9 different cases into the following classes:
\begin{itemize} 
\item Real emdedings exist and are physically interesting. This class contains the embeddings $AdS_{D-1}\subset AdS_{D}$, $dS_{D-1}\subset dS_{D}$, $dS_{D-1}\subset M_{D}$ and $M_{D-1}\subset AdS_{D}$;
\item Real embedding exists but is trivial. This class contains only one case, i.e. $M_{D-1}\subset M_{D}$;
\item Real embeddings exist, but may involve bulk geometries with multiple timelike directions. Not knowing of any physical interpretations of spacetimes with multiple timelike directions, we regard this class of embeddings as containing some physical illness, however if Wick rotations are taken into account, such embeddings may still yield interesting results. This class contains the cases $AdS_{D-1}\subset M_{D}$, $AdS_{D-1}\subset dS_{D}$ and $dS_{D-1}\subset AdS_{D}$;
\item Real embedding is impossible. This class contains only one case, i.e. $M_{D-1}\subset dS_{D}$.
\end{itemize}

We remark here that the above classification makes sense only for codimension 1 embeddings. If more extra dimensions were allowed, some of the impossibilities or illnesses might be resolved. For instance, although the embedding $dS_{D-1} \subset AdS_{D}$ is ill in the sense that it requires more than one timelike directions in the bulk, the embedding $dS_{D-1} \subset AdS_{D+1}$ can be achieved without such illness, e.g. via the chain of embeddings $dS_{D-1} \subset M_{D}\subset AdS_{D+1}$. In the next section we shall consider some of the codimension $>1$ embeddings, but we will focus exclusively on the one step embeddings, i.e. excluding the above mentioned chain-like embeddings.

\section{Codimension $n>1$}

Comparing to the codimension 1 cases the biggest difference of codimension $n>1$ cases lies in that the geometry of the extra dimensional subspace (i.e. that described by the metric $h_{ij}$) is no longer trivial. Thus it is basically impossible to list all types of warped embeddings as we did for the codimension 1 cases, because each types of embeddings listed in (\ref{9cases}) would be subdivided into many different cases according to the choice of geometry $h_{ij}$.  In this section, we shall only take the simplest choices of $h_{ij}$, so that it possesses an $SO(n)$ symmetry, i.e. contains an $(n-1)$-sphere as a subspace. Other possible choices will not be considered. Meanwhile, due to the fact that already in the codimension 1 cases some of the one-step embeddings yield solutions with physical illness or even give rise to no real solution, we shall only consider some of the 9 different types of embeddings listed in (\ref{9cases}), rather than make a complete case by case study. The equations (\ref{munu-D})-(\ref{ij-D}) are now too complicated to be solved directly, so we shall adopt a trial-and-check approach.

\subsection{$AdS_{d} \subset AdS_{D}$ with $n=D-d>1$} \label{cod-higher}

We need to make an assumption for the bulk metric which reduces to the solution (\ref{C1}) in the limit $n=1$. A mathematically viable ansatz for the metric of this kind reads
\begin{equation}
ds^{2}=\left(\frac{k}{\hat{k}}\right)^{2}\cosh^{2}(\hat{k}\rho)\left(g_{\mu\nu}dx^{\mu}dx^{\nu}\right)-d\rho^{2}-A^{2}(\rho)d\Omega_{n-1},
\label{assump}
\end{equation}
where $A(\rho)$ is to be determined by the embedding equations and $d\Omega_{n-1}$ represents the metric of a unit $(n-1)$-sphere. Using the embedding equations we can get
\begin{displaymath}
A(\rho)=\hat{k}^{-1}\sinh(\hat{k}\rho),
\end{displaymath}
i.e. the final bulk metric takes the form
\begin{equation}
ds^{2}=\left(\frac{k}{\hat{k}}\right)^{2}\cosh^{2}(\hat{k}\rho)\left(g_{\mu\nu}dx^{\mu}dx^{\nu}\right)-d\rho^{2}-\hat{k}^{-2}\sinh^{2}(\hat{k}\rho)d\Omega_{n-1}. \label{adsads-n}
\end{equation}

The solution (\ref{adsads-n}) has some interesting features. If we omit the part of the $d$-dimensional submanifold and look only at the extra $n$-dimensional subspace, the metric turns out to be that of an $n$-dimensional cone, with an $(n-1)$-spheric foliation along the $\rho$ axis. The tip of the cone is located at $\rho=0$, where the size of the $(n-1)$-sphere shrinks to zero. That is to say, if we come close to $\rho=0$ in the $D$-dimensional metric (\ref{adsads-n}), the spacetime would appear to be $d$-dimensional. So there is a spontaneous compactification effect near the tip of the cone. This whole picture is very similar to the Klebanov-Strassler model of superstring cosmology \cite{Klebanov}, in which our 4-dimensional universe ``dances at the tip of a pin'' (the end of Klebanov-Strassler throat). However the problem we are considering here is much simpler: we only consider standard General Relativity and no supersymmetry is needed in this picture.

\subsection{$dS_{d} \subset dS_{D}$ with $n=D-d>1$}

Similar considerations can be carried out for the case $dS_{d} \subset dS_{D}$. The solution turns out to be
\begin{equation}
ds^{2}=\left(\frac{k}{\hat{k}}\right)^{2}\cos^{2}(\hat{k}\rho)\left(g_{\mu\nu}dx^{\mu}dx^{\nu}\right)-d\rho^{2}-\hat{k}^{-2}\sin^{2}(\hat{k}\rho)d\Omega_{n-1}. \label{dsds-n}
\end{equation}
Now the geometry of the extra $n$-dimensional subspace is no longer a cone but rather an $n$-sphere. It is interesting to note that there is a see-saw mechanism between the factors $\cos^{2}(\hat{k}\rho)$ in front of the $d$-dimensional submanifold and $\sin^{2}(\hat{k}\rho)$ in front of the $(n-1)$-sphere. Explicitly, near the poles $\hat{k}\rho =0, \pi$ of the $n$-dimensional sphere, the whole spacetime looks $d$-dimensional, while near the equator $\hat{k}\rho =\pi/2$ of the $n$-dimensional sphere, the whole spacetime looks $(n-1)$-dimensional. In either cases the effective dimension of the bulk spacetime is drastically decreased, so this can be understood as another type of spontaneous compactification.

\subsection{$M_{d} \subset M_{D}$ with $n=D-d>1$}

Already in the codimension 1 case the embedding of $M_{D-1}$ into $M_{D}$ has shown its trivialness. No surprises occur at higher codimensions. The only solution we get in the spirit of subsection \ref{cod-higher} is the following,
\begin{equation}
ds^{2}=g_{\mu\nu}dx^{\mu}dx^{\nu}-d\rho^{2}-\rho^{2}d\Omega_{n-1}. \label{mm-n}
\end{equation}
This is nothing but a direct sum of the $d$-dimensional flat submanifold with an $n$-dimensional Euclidean space. Such solutions were used in the construction of (unstable) black branes \cite{GL} in the literature.

\subsection{Other cases}

Though it is our hope to analyse every cases listed in (\ref{9cases}) in detail, it turned out that finding explicit solutions for other types of warped embeddings is very difficult. Even the existence of a real analytic solution is not guaranteed. However, we can check by direct calculations that the following is a valid embedding from a 4-dimensional de Sitter spacetime to a 7-dimensional flat spacetime:
\[
ds^{2}=\frac{1}{5}\left[\Lambda g_{\mu\nu}dx^{\mu}dx^{\nu}-d\rho^{2}-\rho^{2}d\Omega_{2}\right],
\]
but we simply do not know of any generalizations to arbitrary dimensions.

\section{Codimension $n>1$: another branch of solutions}

The solutions described in the last section rely on the assumption (\ref{assump}) or the like, which require that in the limit of $n=1$ the codimension 1 solutions should be recovered. Since the codimension $n>1$ cases have much richer structure, it is natural to ask whether there exist solutions which do not approch the codimension 1 solutions in the limit $n=1$. To answer this question, we can make no reference to the results accumulated in section \ref{S2}. So we take another route: fix the geometry of the extra $n$-dimensional subspace as in the last section and see if there exist other solutions for the warp factors. This seems to be a very strange route to take, but, as it turns out, there indeed exist positive answers to the question. 

\subsection{$AdS_{d} \subset AdS_{D}$ with $n=D-d>1$} 

Instead of (\ref{assump}), we now take the ansatz
\begin{equation}
ds^{2}=B^{2}(\rho)\left(g_{\mu\nu}dx^{\mu}dx^{\nu}\right)-d\rho^{2}-\hat{\kappa}^{-2}\sinh^{2}(\hat{\kappa}\rho)
d\Omega_{n-1}, \label{assump2}
\end{equation}
where $B(\rho)$ is to be determined by the embedding equations. Notice that in writing (\ref{assump2}), we have implicitly used the relation (\ref{lambda-2}). 

Inserting (\ref{assump2}) into the embedding equations we find, after straightforward calculations, that the only solution for $B(\rho)$ is a constant, $
B(\rho)=\left(\frac{\kappa}{\hat{\kappa}}\right)^2$. 
Therefore the bulk metric in this case looks as follows,
\begin{equation}
ds^{2}=\left(\frac{\kappa}{\hat{\kappa}}\right)^2\left(g_{\mu\nu}dx^{\mu}dx^{\nu}\right)-d\rho^{2}-\hat{\kappa}^{-2}\sinh^{2}(\hat{\kappa}\rho)
d\Omega_{n-1}. \label{adsads-n2}
\end{equation}
This solution is quite similar to the one given by (\ref{adsads-n}) but for one thing: the warp factor in front of $g_{\mu\nu}$ becomes constant. Thus the present solution belongs to a completely different branch of solutions. Note that the existence of two different branches of solutions to the embedding equations is not a supprise: similar phenomena has already been reported in \cite{Kinoshita} in the context of Fruend-Robin compactification. The present solution also contains an $n$-dimensional cone metric as a factor, and at the tip $\rho=0$ of the cone, the spacetime becomes $d$-dimensional without the aid of a Fruend-Robin field. So the spontaneous compactification mechanism is still present in (\ref{adsads-n2}), just like in 
(\ref{adsads-n}).

\subsection{$dS_{d} \subset dS_{D}$ with $n=D-d>1$} 

We can make analogous assumptions like (\ref{assump2}) in the case of embedding $dS_{d} \subset dS_{D}$. Following similar steps we get the solution
\begin{equation}
ds^{2}=\left(\frac{\kappa}{\hat{\kappa}}\right)^2\left(g_{\mu\nu}dx^{\mu}dx^{\nu}\right)-d\rho^{2}-\hat{\kappa}^{-2}\sin^{2}(\hat{\kappa}\rho)
d\Omega_{n-1}. \label{dsds-n2}
\end{equation}
This solution also looks like (\ref{dsds-n}) but with the warp factor changed into a constant. Similar to (\ref{dsds-n}), the geometry of the extra $n$-dimensional subspace is an $n$-sphere. But unlike (\ref{dsds-n}), the missing warp factor in front of $g_{\mu\nu}$ makes the see-saw mechanism in the bulk metric (\ref{dsds-n2}) disappear. Accordingly, the bulk metric will still reduce to a $d$-dimensional Einstein manifold at the poles $\hat{\kappa}\rho=0,\pi$ of the $n$-sphere, but it will not reduce to an $(n-1)$-sphere at the equator $\hat{\kappa}\rho=\pi/2$ of the $n$-sphere.

\subsection{Other cases}

For the embedding $M_{d}\subset M_{D}$, there is no difference between the two parametrization schemes (\ref{lambda}) and  (\ref{lambda-2}). Thus we should get the same answer as in (\ref{mm-n}) even if we start from an ansatz like (\ref{assump2}) and taking the limit $\hat{\kappa}\rightarrow 0$. 

We don't have explicit solutions to other cases like $dS_{d} \subset M_{D}$ or $M_{d} \subset AdS_{D}$ at hand.
However we expect that there should be some interesting solutions to such embeddings. We wish to make further study on such cases in later works.

\section{Discussions} 

Embeddings between Einstein manifolds of different dimensions have found significant applications in various areas in gravitational physics. In this article, we presented some explicit results of such warped embeddings, focusing on the 
codimension $n>1$ cases. Now let us make some discussions on the potential applications of the codimension $n>1$ solutions obtained in this article. 

The first potential application is on a possible spontaneous compactification mechanism. As we have already noticed earlier, for the embeddings $AdS_{d} \subset AdS_{D}$, both branches of solutions share the common feature that the bulk metric reduces to that of the $d$-dimensional submanifold at the tips of the extra $n$-dimensional cone, and for the embeddings $dS_{d} \subset dS_{D}$, both branches of solutions share a similar feature: the bulk metric reduces to that of the $d$-dimensional submanifold at the poles of the $n$-sphere.
Therefore it is interesting to ask whether there exists a mechanism such that it makes the $d$-dimensional submanifolds prefer to stay at the special positions mentioned above (tips of the cone or poles of the $n$-spheres). Possible choices for the mechanisms include introducing tension terms for the $d$-dimensional submanifolds or adding matter to these submanifolds. 

Another potential application is in the studies of higher dimensional black holes and black branes. The trivial embedding (\ref{mm-n}) and its codimension 1 analogue has already been used in the studies of black branes and strings and was shown to give rise to black brane/string instabilities \cite{GL}. In the last few years many five-dimensional black hole solutions with horizon topology $S^2 \times S^1$ (i.e. ``black rings'') were found in flat spacetime (i.e. with vanishing cosmological constant). Among these black rings the so-called large black rings are thermodynamically more favored and are believed to be more stable. Attempts in finding black ring solutions in the presence of a cosmological constant have since been made consecutively. However no analytic, singularity-free solutions of this kind have been found to this date.  
We postulate that the embeddings considered in this article may be helpful in constructing higher dimensional black ring solutions with nonvanishing cosmological constant. However the application of codimension 1 embeddings in constructing black ring solutions was a failure \cite{Chu}: the resulting spacetime contains naked singularity and hence is not a black ring solution in the usual sense. The inclusion of more extra dimensions may help to resolve the naked singularity problem and give rise to physically acceptable black ring solutions in higher dimensions. We shall attempt to make more explorations in this direction in the future.

\section*{Acknowledgement}

This work was initiated when both authors were visiting KITPC.  L.Z. is supported by the Natural Science Foundation of China (NSFC) through grant No. 10875059.

\providecommand{\href}[2]{#2}\begingroup\raggedright\endgroup


\begin{thebibliography}{10}

\bibitem{Leblond} F. Leblond, R. Myers, D.J. Winters, Consistency conditions for braneworlds in arbitrary dimensions, JHEP 0107 (2001) 031 [arXiv: hep-th/0106140];

\bibitem{Pope} I. Y. Park, C. N. Pope and A. Sadrzadeh, AdS braneworld Kaluza-Klein reduction,
Class. Quant. Grav. 19 (2002) 6237 [arXiv:hep-th/0110238];

\bibitem{Randall1} L. Randall, R. Sundrum, Large Mass Hierarchy from a Small Extra Dimension, Phys. Rev. Lett. 83, No.17 (1999) 3370;

\bibitem{Randall2} L. Randall and R. Sundrum, An alternative to compactification, Phys. Rev. Lett. 83,  
4690 (1999) [arXiv:hep-th/9906064];

\bibitem{Chu} C.-S. Chu, S.-H. Dai, On Black Ring with a Positive Cosmological Constant, Phys. Rev. D75 (2006) 064016 [arXiv:hep-th/0611325];

\bibitem{Zhao} L. Zhao, K. Niu, B. S. Xia, Y. L. Dou, J. Ren, 
Non-uniform Black Strings with Schwarzschild-(Anti-)de Sitter
Foliation, Class. Quant. Grav. 24 (2007) 4587 [arXiv:hep-th/0703195];

\bibitem{Klebanov} I. R. Klebanov and M. J. Strassler, Supergravity and a confining gauge theory, 
Duality cascades and $\chi$SB-resolution of naked singularities, JHEP 0008, 052 (2000) [arXiv:hep-th/0007191]; 

\bibitem{GL} R. Gregory, R. Laflamme, Black strings and p-branes are unstable, Phys.Rev.Lett. 70 (1993) 2837, [arXiv:hep-th/9301052];

\bibitem{Kinoshita} S. Kinoshita, New branch of Kaluza-Klein compactification, Phys. Rev. D76 (2007) 124003 [arXiv:0710.0707].


\end{thebibliography}
\end{document}